\documentclass[prl, twocolumn, amsmath, amssymb, superscriptaddress]{revtex4-2}

\usepackage{graphicx, braket, xcolor}

\usepackage[pdftex,colorlinks=true,citecolor=black,urlcolor=black,linkcolor=black]{hyperref}

\usepackage[inkscapelatex=false]{svg}

\newcommand{\+}{^{\dagger}}
\newcommand{\ha}{\hat{a}}

\newcommand{\Rev}[1]{{\color{black} #1}}

\usepackage[mathscr]{euscript}

\begin{document}
\title{Exploiting nonclassical motion of a trapped ion crystal for quantum-enhanced metrology of global and differential spin rotations}

\author{R.~J. Lewis-Swan}
\affiliation{Homer L. Dodge Department of Physics and Astronomy, The University of Oklahoma, Norman, Oklahoma 73019, USA}
\affiliation{Center for Quantum Research and Technology, The University of Oklahoma, Norman, Oklahoma 73019, USA}
\author{J.~C. Zu\~{n}iga Castro}
\affiliation{Homer L. Dodge Department of Physics and Astronomy, The University of Oklahoma, Norman, Oklahoma 73019, USA}
\affiliation{Center for Quantum Research and Technology, The University of Oklahoma, Norman, Oklahoma 73019, USA}
\author{D. Barberena}
\affiliation{JILA, NIST, and Department of Physics, University of Colorado, Boulder, CO, USA}
\affiliation{Center for Theory of Quantum Matter, University of Colorado, Boulder, CO, USA}
\author{A.~M. Rey}
\affiliation{JILA, NIST, and Department of Physics, University of Colorado, Boulder, CO, USA}
\affiliation{Center for Theory of Quantum Matter, University of Colorado, Boulder, CO, USA}

\begin{abstract}
        We theoretically investigate prospects for the creation of nonclassical spin states in trapped ion arrays by coupling to a squeezed state of the collective motion of the ions. The correlations of the generated spin states can be tailored for quantum-enhanced sensing of global or differential rotations of sub-ensembles of the spins by working with specific vibrational modes of the ion array. We propose a pair of protocols to utilize the generated states \Rev{and demonstrate their viability even for small systems, while assessing limitations imposed by spin-motion entanglement} and technical noise. Our work suggests new opportunities for the preparation of many-body states with tailored correlations for quantum-enhanced metrology in spin-boson systems.
\end{abstract}

\maketitle 

\noindent\emph{Introduction:} Preparing entangled atomic states is a continuing challenge for the realization of quantum-enhanced sensors. A strong focus has been on generating collective states as these are optimal for applications including atomic clocks and interferometers \cite{Pezze_2018,szigeti2021review}. However, recent work has drawn attention to the engineering of states with distributed entanglement and correlations for multi-parameter estimation or quantum sensors with enhanced spatial resolution \cite{Jing_2019,Gessner2020,fadel2022splitspin,kasevich2022distributed,Cooper2022graph,zoller2023multiparameter,sinatra2023multiparameter,braun2023noncommuting}. 

One way to prepare collective entangled states is to realize global spin-spin interactions mediated by a common bosonic mode uniformly coupled to a spin ensemble, such as in trapped ion \cite{Britton2012ising,yao2021ionreview} and cavity-QED \cite{Norcia_2018,davis2019spin1,leroux2010squeezing} platforms. In the former, squeezed states -- featuring a reduction in quantum projection noise for specific observables -- have been realized in both one-dimensional ($1$D) \cite{muleady2023squeezing} and two-dimensional ($2$D) ion arrays \cite{Bohnet2016}. 
However, the requirement to operate in a far-detuned regime can lead to 
challenges such as slow timescales for entangling dynamics relative to intrinsic decoherence, and spurious couplings to other boson modes that lead to a reduction in the effective range of spin-spin interactions. Both issues limit the amount of squeezing that can be generated. 

Concurrently, the trapped ion community has made strives forward in the coherent control of the quantized vibrational motion of the ions for quantum information processing, simulation and logic spectroscopy \cite{meekhof1996nonclassical,monroe1996cat,schmidtQLS_2005,haffner2008ions,Home2019motionalqubit,Burd2019,Burd2021,leibfried2022beamsplitter}. In this light, we investigate the feasibility of creating entangled spin states through coherent transfer of squeezed fluctuations from the motional to the spin degree of freedom, which builds on early ideas to generate squeezing with trapped ions \cite{wineland1992squeezing,wineland1994squeezing} and demonstrated in atom-light systems \cite{Kuzmich1997squeezing,polzik1999squeezing}. 

\Rev{Our proposal to use a resonant spin-boson coupling with a single mode leads to states featuring squeezing of a fixed spin quadrature and a coherent transfer time  that is independent of the degree of the initial squeezing. Moreover, we show that spatially inhomogeneous spin-boson couplings can be used to create spin states with enhanced sensitivity to differential rotations between two parts of the array. The latter capability can have potential applications for clock comparisons \cite{bhuvanesh_2023_TMS,robinson_2024_comparison}, gravitational redshift measurements \cite{bothwell_2022_redshift} or magnetometry \cite{ruster_2017_magnetometry}.} \Rev{The generated states can enable quantum-enhanced Ramsey interferometry even for small numbers of ions, relevant for near-term experiments, with performance that is fundamentally constrained by the buildup of spin-motion entanglement. To overcome this issue, and exploit the generated squeezing without the need for site-resolved  measurements, we propose an interaction-based readout (IBR) protocol based on time-reversed dynamics} \cite{Davis_2016,Nolan_2017,gilmore2021quantum,Colombo2022timereversal,agarwal2022amplification} that disentangles the degrees of freedom and requires only global manipulations and measurements of the spins.

\noindent\emph{Spin-boson toolkit:} We consider a linear chain of $N$ ions with $m=1,2,...,N$ axial phonon modes with harmonic frequencies $\omega_m$ and associated bosonic creation (annihilation) operators $\hat{a}^{\dagger}_m$ ($\hat{a}_m$). We focus on axial motion in $1$D for simplicity, but it is straightforward to extend our analysis to radial modes or higher dimensional arrays. A spin-$1/2$ is encoded in a pair of internal states $\ket{\downarrow}$ and $\ket{\uparrow}$ of each ion. State-of-the-art trapped ion quantum simulators provide a toolbox of operations to manipulate and couple spin and motion.

Global spin rotations are realized by driving the qubits with optical or microwave fields. The phonons can be manipulated by, e.g., modulating the confining potential of the ion chain or additional electric fields, to realize single-mode squeezing \cite{wineland1990amplification,FossFeig2019amplification,Burd2019,Burd2021} or a coherent coupling between pairs of modes \cite{leibfried2022beamsplitter}. The former is described by the unitary operation $\hat{S}(\zeta) = e^{\frac{1}{2}\left( \zeta^*\hat{a}_m^2 - \zeta \hat{a}_m^{\dagger 2} \right)}$ where $\zeta = r e^{i\phi}$ is the squeezing parameter with strength $r$ and phase $\phi$. Squeezing reduces the fluctuations along one bosonic quadrature at the expense of increased fluctuations in an orthogonal quadrature. For example, for $\phi = 0$ squeezing transforms $\langle (\Delta \hat{X})^2 \rangle = \langle (\hat{X} - \langle\hat{X}\rangle)^2 \rangle \to e^{-2r}\langle (\Delta \hat{X})^2 \rangle$ and $\langle (\Delta \hat{Y})^2 \rangle \to e^{2r}\langle (\Delta \hat{Y})^2 \rangle$ where $\hat{X} = \hat{a}+\hat{a}^{\dagger}$ and $\hat{Y} = i(\hat{a}^{\dagger} - \hat{a})$. 
The coherent coupling of phonon modes $m$ and $n$ is described by the unitary operation $\hat{U}_{\mathrm{bs}}(\kappa_{mn}) = e^{i\kappa_{mn}(\hat{a}^{\dagger}_m\hat{a}_n + \hat{a}^{\dagger}_n\hat{a}_m)/2}$ where setting $\kappa_{mn} = \pi$ realizes a perfect swap of the quantum states of each mode.

\begin{figure}
    \centering
    \includegraphics[width=8cm]{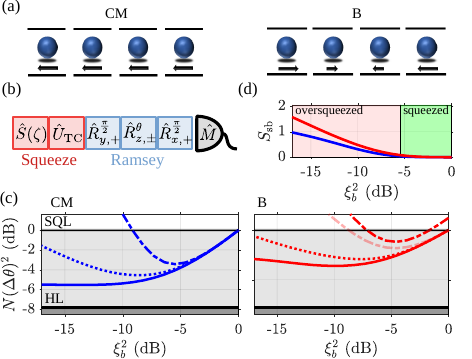}
    \caption{(a) Example modes of a $1$D chain. Arrows qualitatively indicate the ion participation in each mode. (b) Sequence for NR protocol. 
    (c) Metrological gain $N(\Delta\theta)^2$ as a function of boson squeezing $\xi^2_b$. The performance achieved with measurements of $\hat{M} = \hat{S}_{z,+}$ (CM, dot-dashed line), $\hat{\mathscr{S}}_z$ (B, dot-dashed line) and $\hat{S}_{z,-}$ (B, faded dot-dashed line) is compared to the QCRB given by the QFI of the full system, $N(\Delta\theta)^2 = N/F_Q$ (solid lines). \Rev{We also compare against the spin-only QFI via $N/F_{Q,s}$ (dotted lines). (d) Spin-boson entanglement $S_{\mathrm{s}\mathrm{b}}$ [colors same as (c)]. Panels (c) and (d) use $N=6$.} 
    }
    \label{fig:Ramsey}
\end{figure}

Spin-motion coupling can be realized by driving a red sideband transition, described by the (inhomogeneous) Tavis-Cummings Hamiltonian \cite{retamal2007TC,reznik2007taviscummings,SM},
\begin{equation}\label{eqn:HTC}
    H_{\mathrm{TC},m} = \sum_{j=1}^N g_{jm} \left( \hat{a}^{\dagger}_m \hat{\sigma}^{-}_j + \hat{a}_m\hat{\sigma}^+_j \right) .
\end{equation}
The coupling $g_{jm}$ is determined by the participation of the $j$th ion in the $m$th mode. In this work we focus on the center-of-mass (CM) and breathing (B) modes [see Fig.~\ref{fig:Ramsey}(a)]. The former couples uniformly, $g_{jm} = g_0/\sqrt{N}$ with $g_0$ the characteristic spin-boson coupling strength, while the latter is given by the inhomogeneous coupling $g_{jm} = g_0 u_j/\sqrt{\sum_j u_j^2}$ with $u_j$ the equilibrium position of each ion in the crystal (the origin is chosen to lay at the center of chain) \cite{James1998}.

\noindent{\emph{NR protocol:}} We first investigate a noise-reduction (NR) protocol [Fig.~\ref{fig:Ramsey}(b)] that generates spin squeezing for a Ramsey sequence. The ions are cooled into the motional ground-state of the CM (B) mode and the qubits are prepared uniformly in $\ket{\downarrow}$. 
The vacuum fluctuations of the CM (B) mode are then squeezed, such that the quadrature variances are $\langle (\Delta\hat{X})^2 \rangle = e^{-2r}$ and $\langle (\Delta\hat{Y})^2 \rangle = e^{2r}$ with $\phi$ chosen to be zero.
Next, a Tavis-Cummings interaction is applied for a time $t_{\pi} = \pi/(2g_0)$ [denoted by $\hat{U}_{\mathrm{TC}}$ in Fig.~\ref{fig:Ramsey}(b)], ideally leading to a coherent exchange of the fluctuations between the phonons and the spin ensemble \cite{wineland1994squeezing,sanders2003squeezing} and thus preparing a squeezed spin state.

\Rev{Insight into the underlying mechanism can be found in the large $N$ limit} by applying a Holstein-Primakoff transformation to the collective raising (lowering) operators, $\sum_{j=1}^N \sigma^+_j \to \sqrt{N}\hat{b}^{\dagger}$ ($\sum_{j=1}^N \sigma^-_j \to \sqrt{N}\hat{b}$) \cite{SM}.
Equation (\ref{eqn:HTC}) then becomes $\hat{H}_{\mathrm{TC},\mathrm{CM}}=g_0(\hat{a}^{\dagger}_{\mathrm{CM}}\hat{b}+\hat{a}_{\mathrm{CM}}\hat{b}^{\dagger})$, which describes a beam-splitter between two bosonic modes. Evolving under $\hat{H}_{\mathrm{TC},\mathrm{CM}}$ for $t_{\pi}$ thus coherently exchanges the states of the spin and motion. In particular, the collective spin fluctuations after the TC interaction are given by $\langle (\Delta \hat{S}_{y,+})^2 \rangle = Ne^{-2r}/4$ and $\langle (\Delta \hat{S}_{x,+})^2 \rangle = Ne^{2r}/4$ where $\hat{S}_{\alpha,+} = \frac{1}{2}\sum_{j=1}^N \hat{\sigma}^{\alpha}_j$ for $\alpha =x,y,z$. Spin squeezing is witnessed by $\xi^2_s < 1$ where $\xi^2_{s} = N \langle (\Delta \hat{S}_{y,+})^2 \rangle/\vert \langle \hat{\boldsymbol{S}} \rangle \vert^2$ and $\hat{\boldsymbol{S}} = (\hat{S}_{x,+},\hat{S}_{y,+},\hat{S}_{z,+})$ \cite{wineland1992squeezing}. \Rev{More generally, the squeezed spin quadrature can be precisely controlled by varying $\phi\neq0$ or the phase of the couplings $g_{jm}$.} Similar results are expected for the B mode, though it features squeezed fluctuations of the weighted spin operators $\hat{\mathscr{S}}_{\alpha} = \frac{\sqrt{N}}{2g_0}\sum_{j=1}^N g_{j,\mathrm{B}} \hat{\sigma}^{\alpha}_j$ for $\alpha =x,y,z$ such that $\langle (\Delta \hat{\mathscr{S}}_{y})^2 \rangle \approx Ne^{-2r}/4$ and $\langle (\Delta \hat{\mathscr{S}}_{x})^2\rangle \approx Ne^{2r}/4$. 

The spin squeezing can subsequently be exploited for metrology using a Ramsey sequence composed of: i) a global $\pi/2$ qubit rotation about $\hat{y}$, ii) an interrogation period where a phase $\theta$ is imprinted by a collective (differential) rotation about $\hat{z}$ described by $\hat{R}^{\theta}_{z,+} = e^{-i\theta\hat{S}_{z,+}}$ [$\hat{R}^{\theta}_{z,-} = e^{-i\theta\hat{S}_{z,-}}$ where $\hat{S}_{z,-} = \frac{1}{2}\left(\sum_{j=1}^{N/2} \hat{\sigma}^z_j - \sum_{j=N/2+1}^N \hat{\sigma}^z_j\right)$], and iii) a global $\pi/2$ qubit rotation about $\hat{x}$. 
\Rev{The parameter $\theta$ is estimated by measuring the spin-projections $\hat{M} = \hat{S}_{z,+}$ (CM) and $\hat{M} = \hat{\mathscr{S}}_z$ or $\hat{S}_{z,-}$ (B), with associated sensitivity characterized by the metrological gain $N(\Delta\theta)^2 = N\langle (\Delta \hat{M})^2 \rangle/\partial_{\theta} \langle \hat{M} \rangle \vert^2$ (equivalent to the spin squeezing $\xi^2_s$ of the prepared state).} Sensitivity surpassing the standard quantum limit corresponds to $N(\Delta\theta)^2 < 1$ with a lower bound $(\Delta\theta)^2 \geq 1/N^2$ given by the Heisenberg limit.

\noindent{\emph{Limitations of the NR protocol:}} We assess the role of finite size effects and the inhomogeneous spin-boson couplings for the B mode by numerically simulating the NR protocol and show the results in Fig.~\ref{fig:Ramsey}(b)-(d). \Rev{All results are obtained by numerical integration of the Schrodinger equation unless stated otherwise.}

\begin{figure}
    \centering
    \includegraphics[width=8cm]{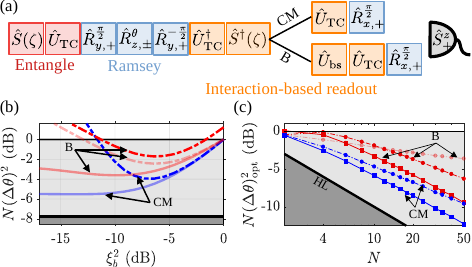}
    \caption{(a) Sequence for SA protocol. The final steps of the IBR depend on the coupled mode and measurement. (b) Metrological gain as a function of boson squeezing $\xi^2_b$ using $\hat{M} = \hat{S}_{z,+}$ (CM and B, dot-dashed blue and red lines) and $\hat{S}_{z,-}$ (B, faded dot-dashed red line). We also plot the QCRB $N/F_Q$ (faded blue and red lines). \Rev{Calculations use $N=6$.} (c) Scaling of optimal metrological gain with ion number $N$ for NR (circles) and SA (squares) protocols for the CM (blue) and B (red) modes. We distinguish $\hat{S}_{z,-}$ (faded) and $\hat{\mathscr{S}}_z$ (darker) measurements for the B mode NR protocol. \Rev{Data for this panel is obtained using a truncated Wigner approximation \cite{asier2017dtwa,SM}.}}
    \label{fig:TimeReversal}
\end{figure}

Figure \ref{fig:Ramsey}(c) shows the metrological gain (blue lines) as a function of the boson squeezing $\xi^2_b = e^{-2r}$ for \Rev{$N=6$} qubits. 
When coupling to the CM mode we find near perfect exchange of fluctuations, $\xi^2_s \approx \xi^2_b$, for ``weak'' boson squeezing, before an optimal spin squeezing $\xi^2_{s,\mathrm{opt}} \approx 3.5$~dB is reached at $\xi^2_{b,\mathrm{opt}} \approx 5.5~$dB. For $\xi^2_b < \xi^2_{b,\mathrm{opt}}$ there is an \emph{oversqueezed} regime where spin squeezing is quickly lost. The observation of an optimal $\xi^2_{s,\mathrm{opt}}$ is consistent with the twofold expectations that: i) an ensemble of $N$ qubits can only support a finite amount of squeezing (most stringently, $\xi^2_s \geq 1/N$), and ii) the interpretation of the TC interaction as an effective beam-splitter is only valid for large $N$ or correspondingly moderate boson squeezing. 
For both aspects, one should be cognizant that the bosonic fluctuations span a flat $2$D phase space defined by the quadratures $(X,Y)$, whereas the spin fluctuations lie on the curved surface of the collective Bloch sphere with axes $(S_x,S_y,S_z)$.
The large $N$ limit approximates the spin fluctuations to occupy only the tangential $S_x$-$S_y$ plane perpendicular to the initial polarization of the spins along $-\hat{z}$. 
For modest boson squeezing ($\xi^2_b > \xi^2_{b,\mathrm{opt}}$) this plane is sufficient to describe the squeezed noise exchanged onto the spins, but for large boson squeezing ($\xi^2_b < \xi^2_{b,\mathrm{opt}}$) it fails as the anti-squeezed projection noise probes the curved surface of the Bloch sphere. 
To illustrate this, we extend our study to larger systems [see Fig.~\ref{fig:TimeReversal}(c)] and find that the optimal spin squeezing asymptotically scales as \Rev{$\xi^2_{s,\mathrm{opt}} \propto N^{-0.68\pm 0.02}$ (uncertainty indicates $95$\% confidence interval including fitting error)} \cite{SM}. This is approximately identical to what can be obtained with one-axis twisting \cite{Kitagawa1993} and we comment on this shortly.

Similar results are observed for the B mode, although the \Rev{optimal gain, $N(\Delta\theta)^2_{\mathrm{opt}} \approx 1.2$~dB at $\xi^2_{b,\mathrm{opt}} \approx 4.5$~dB, is slightly worse. This is primarily due to $\hat{\mathscr{S}}_z$ not quite being the optimal observable to estimate $\theta$, as evidenced by $N(\Delta\theta)^2 > 1$ as $\xi^2_b \to 1$ \cite{SM}. Given that} measurement of the weighted spin-projection requires detailed knowledge of the couplings $g_{j,\mathrm{B}}$, \Rev{we also consider the metrological performance for a measurement of the simpler differential magnetization $\hat{S}_{z,-}$, which accounts only for the alternating sign of the B mode coupling across the ion chain [see Fig.~\ref{fig:Ramsey}(a)]. For $N = 6$ this actually leads to slightly superior performance [$N(\Delta\theta)^2_{\mathrm{opt}} \approx 1.8$~dB]. However, this quickly changes with system size [see Fig.~\ref{fig:TimeReversal}(c)] and $\hat{\mathscr{S}}_z$ becomes preferable: $N(\Delta\theta)^2_{\mathrm{opt}} \propto N^{-0.61\pm0.03}$ and $N^{-0.22\pm0.01}$ for the differential and weighted observables, respectively \cite{SM}.}

To further characterize the metrological potential of the prepared probe state we compute the quantum Fisher information (QFI) $F_Q = 4\langle (\Delta \hat{S}_{x,\pm})^2 \rangle$, which constrains the best sensitivity (optimized over all measurements) by the quantum Cramer-Rao bound (QCRB) $(\Delta\theta)^2 \geq F_Q^{-1}$. We plot the optimal metrological gain $NF_Q^{-1}$ in both panels of Fig.~\ref{fig:Ramsey}(c) (red lines). The QFI predicts a significantly enhanced metrological gain relative to spin squeezing for $\xi^2_b < \xi^2_{b,\mathrm{opt}}$ and saturates to $NF_Q^{-1} \approx N/2$ for large boson squeezing.

The \emph{oversqueezed} regime featuring large QFI but poor squeezing as a result of the curved Bloch sphere is reminiscent of one-axis twisting protocols in collective spin systems \cite{Kitagawa1993,Davis_2016}. 
For these, measurements of higher-order observables \cite{Gessner2019Nonlinear,gessner2022nonlinear} or counting statistics \cite{Strobel2014,Bohnet2016} can be used to approach the QCRB. In contrast, oversqueezing in our protocol is associated with spin-boson entanglement. 
Figure \ref{fig:Ramsey}(d) shows the Renyi entanglement entropy $S_{\mathrm{sb}} = -\mathrm{log}[\mathrm{Tr}(\hat{\rho}_s^2)]$, where $\hat{\rho}_s = \mathrm{Tr}_{\mathrm{ph}}[\hat{\rho}]$ is the reduced density matrix of the spins. 
While $S_{\mathrm{sb}}$ is vanishingly small in the squeezed regime ($\xi^2_b > \xi^2_{b,\mathrm{opt}}$), it grows appreciably in the oversqueezed regime ($\xi^2_b < \xi^2_{b,\mathrm{opt}}$) and the entanglement leads to excess projection noise in the reduced spin subsystem (as it becomes mixed), thereby limiting the sensitivity attainable with spin measurements. For the CM case, we illustrate this by a calculation of the QFI of the spin subsystem, $F_{Q,s}(\rho_s)$ \cite{SM} [magenta dotted line in panel (b)], satisfying $F_{Q,s} \leq F_Q$ and quantifying the metrological potential of the prepared state when constrained to spin measurements. We observe that $NF_{Q,s}^{-1}$ is appreciably worse than $NF_Q^{-1}$ in the oversqueezed regime, implying that joint measurements of the spins and bosons are required to saturate the QCRB.

\noindent\emph{SA protocol:} To exploit the oversqueezed regime we propose a signal amplification (SA) protocol based on IBR. An example sequence is shown in Fig.~\ref{fig:TimeReversal}(a): We supplement the NR protocol by a time-reversal sequence where the TC interaction and boson squeezing operation are undone (achieved by flipping the sign of the respective Hamiltonian through single qubit manipulations and/or jumping the phase of applied lasers and electric fields) and a final mode-dependent readout step.

In the large $N$ limit, the time-reversal sequence maps the rotation of the complex, entangled probe state to a simple, disentangled product state. Specifically, undoing the TC interaction transforms the spin rotation into an effective coherent displacement of the phonon mode, $\hat{U}_{\mathrm{TC}} \hat{R}^{\frac{\pi}{2}}_{y,+} \hat{R}^{\theta}_{z,\pm} \hat{R}^{\frac{\pi}{2}}_{y,+} \hat{U}_{\mathrm{TC}}^{\dagger} \equiv \hat{D}(\sqrt{N}\theta)$ where $\hat{D}(\alpha) = e^{i\alpha\hat{Y}}$. This displacement is amplified by the squeezing-unsqueezing of the phonon mode according to $\hat{S}^{\dagger}(\zeta)\hat{D}(\sqrt{N}\theta)\hat{S}(\zeta) \equiv \hat{D}(e^{r}\sqrt{N}\theta)$ \cite{Ge2019,FossFeig2019amplification,Burd2019}. After time-reversal, $\theta$ is encoded solely in the displacement of the phonon mode, which is typically not amenable to direct detection. Thus, we use an additional TC interaction to transform the phonon displacement to a rotation of the spin ensemble by an angle $e^r\theta$ about $-\hat{z}$, which can be characterized by, e.g., a simple measurement of the collective magnetization via fluorescence.

The protocol for the B mode is understood analogously, with the final TC interaction followed by a measurement of $\hat{\mathscr{S}}_z$ \Rev{or $\hat{S}_{z,-}$}. However, one could also add a beam-splitter operation $\hat{U}_{\mathrm{bs}}$ that couples the CM and B modes after the time-reversal sequence \cite{leibfried2022beamsplitter} [see lower sequence in Fig.~\ref{fig:TimeReversal}(a)], which interchanges the state of the displaced B mode with the unused CM mode. The rotation angle $\theta$ is inferred from the collective magnetization after a final TC interaction. This alternative sequence enables sensing of differential rotations with no requirement for single-ion resolution or manipulation.

Figure \ref{fig:TimeReversal}(b) shows the metrological gain achieved with the SA protocol as a function of boson squeezing for an example of \Rev{$N=6$} ions. For the B mode, the same gain is obtained whether $\hat{\mathscr{S}}_z$ or $\hat{S}_{z,+}$ (after the coupling of the CM and B modes) is measured and thus we only plot the former in Figs.~\ref{fig:TimeReversal}(b) and (c). We find the optimal metrological gain is \Rev{$N(\Delta\theta)^2_{\mathrm{opt}} = 4$~dB and $2$~dB for $\xi^2_b \approx 6$~dB using the CM and B modes, respectively.} Additionally, we find $F_{Q,s} = F_Q$ for all $\xi^2_b$ (see Ref.~\cite{SM}) and point out that the QCRB (faded blue and red lines in Fig.~\ref{fig:TimeReversal}(b)) could be saturated by a more demanding measurement of the projection onto the initial spin state \cite{Macri_2016,Martin2017_OTOC} or spin counting statistics \cite{SM}. The enhancement provided by the SA protocol is emphasized with increasing system size.
Figure \ref{fig:TimeReversal}(c) shows the the optimal metrological gain as a function of $N$ and we find \Rev{$N(\Delta\theta)^2_{\mathrm{opt}} \propto N^{-0.87\pm0.03}$ and $N^{-0.77\pm0.06}$} for coupling to the CM and B modes, respectively \cite{SM}.

\noindent\emph{Decoherence and noise:} Various technical factors contribute to the performance of our protocols in practice. Imperfect cooling of the targeted normal mode leads to a thermal occupation of $\bar{n}$ phonons and thus excess motional fluctuations before squeezing is applied. This excess noise is inherited by the spin ensemble and also exacerbates finite size effects during the TC interaction. Overall, we predict a degradation in the metrological gain by a factor of $(2\bar{n}+1)^2$ relative to the ideal case \cite{SM}. State-of-the-art trapped ion experiments routinely cool normal modes to near vacuum ($\bar{n} \ll 1$) \cite{Elena2019EIT,Monroe2020EIT,Roos_2Dcrystal_2023}.

\Rev{Damping or heating} of the normal modes at a characteristic rate $\kappa$ during the protocols can have two relevant effects. First, we require $g_0 \gg \kappa$ so that the TC interaction is much faster than the relevant motional decoherence, which would otherwise degrade the squeezing transferred to the spin state and the efficacy of the IBR. 
Simultaneously, $g_0$ (and thus $\kappa$) should be small compared to the relevant frequency spacing of the normal modes near the CM or B modes, so that Eq.~(\ref{eqn:HTC}) is valid. 
This may be a consideration for larger $1$D chains with closely spaced axial modes but we emphasize that our proposal can be extended to radial modes or higher-dimensional arrays. Secondly, the SA protocol may be sensitive to motional decoherence during long phase interrogation periods if operating with states featuring spin-boson entanglement. Spin decoherence can also be relevant, although our protocol occurs on a fixed timescale set by $t_{\pi} = 2\pi/g_0$ whereas squeezing via spin-spin interactions can be more susceptible to decoherence due to intrinsically slower timescales that increase with system size \cite{Bohnet2016,muleady2023squeezing,SM}. 
In addition, the impact of off-resonant light scattering on the sideband protocols that we discuss can scale more favorably with the detuning from relevant internal states than protocols based on spin-spin interactions \cite{bollinger2023spontaneousemission,SM}.

\begin{figure}
    \centering
    \includegraphics[width=8cm]{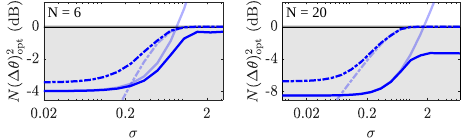}
    \caption{Optimal metrological gain as a function of phase noise magnitude $\sigma$ for NR (dot-dashed lines) and SA (solid lines) protocols \Rev{with $N=6$ and $N=20$} for CM mode. 
    Faded lines are approximate analytic predictions \cite{SM}. 
    }
    \label{fig:Noise}
\end{figure}


\Rev{A lack of phase coherence between the independent fields driving the bosonic squeezing and spin-motion coupling lead to effective shot-to-shot fluctuations of the squeezing angle $\phi$ \cite{SM}. In the large $N$ limit, the NR protocol with the CM mode is limited to $N(\Delta\theta)^2_{\mathrm{opt}} \approx 2\sigma$ where $\sigma \ll 1$ is the rms fluctuation of the squeezing angle, and squeezing is lost entirely for $\sigma \gg 1$. In contrast, the SA protocol yields $N(\Delta\theta)^2 \approx e^{-2r}(1 + 2\sigma^2)$ for $\sigma\ll 1$ and $N(\Delta\theta)^2 \approx 4 e^{-2r}$ for $\sigma \gg 1$ \cite{SM}. Thus, we only require independent stability of the boson and spin operations to retain a quantum-enhancement. Numerical calculations for $N = 6$ and $N = 20$ ions are shown in Fig.~\ref{fig:Noise}. While finite size effects are stronger in the former case, our results are still qualitatively consistent with the large $N$ predictions. Similar robusteness is found for the B mode \cite{SM}.}

\noindent\emph{Summary and outlook:} Our work suggests new opportunities for many-body state preparation and sensing with trapped ions by exploiting the available control over both spin and motion. Arrays in $2$D and $3$D can provide further diversity of normal modes for preparing spin states with complex spatially-structured correlations, while bespoke modes can be created by trapping multiple species \cite{Sosnova2021modes} or additional tweezer potentials \cite{Arghavan2021tweezers}. Our results complement recent studies of multi-parameter estimation in collective spin systems \cite{Gessner2020,fadel2022splitspin,sinatra2023multiparameter,Jing_2019,Kaubruegger2023}, with the distinction that by exploiting the natural structure of collective motion in ion crystals we only require global control and imaging of the qubits. 

\begin{acknowledgments}
\emph{Acknowledgments --} We acknowledge stimulating discussions with J.~J. Bollinger, D.~T.~C Allcock and A.~L. Carter.  R.~J. L-S. and J.~Z acknowledge support by NSF through Grant No. PHY-2110052 and by the U.S. Department of Energy, Office of Science, Advanced Scientific Computing Research, Exploratory Research for Extreme-Scale Science program. A.M.~R. and D.~B. acknowledge support from ARO W911NF-19-1-0210, NSF JILA-PFC PHY-2317149, NSF QLCI-2016244, the DOE Quantum Systems Accelerator (QSA) grant, The Simons Foundation and by NIST. The computing for this project was performed at the OU Supercomputing Center for Education \& Research (OSCER) at the University of Oklahoma (OU).
\end{acknowledgments}

\bibliography{library}

\newpage 

\onecolumngrid
\vspace{\columnsep}
\begin{center}
\textbf{\large Supplemental Material: Exploiting nonclassical motion of a trapped ion crystal for quantum-enhanced metrology of global and differential spin rotations}
\end{center}
\vspace{\columnsep}
\twocolumngrid

\setcounter{equation}{0}
\setcounter{figure}{0}
\setcounter{table}{0}
\makeatletter
\renewcommand{\theequation}{S\arabic{equation}}
\renewcommand{\thefigure}{S\arabic{figure}}

\section{Realization of Tavis-Cummings interaction \label{sec:TCdetails}}
\Rev{The general Hamiltonian describing the collective sideband interaction between the spins and phonons of an ion array can be written as~\cite{reznik2007taviscummings}, 
\begin{equation}
    H_{\mathrm{TC}, m} = \sum_{j = 1}^{N}{g_{jm}(\hat{a}^{\dagger}_m\hat{\sigma}_j^-e^{-i\phi_j} + \hat{a}_m \hat{\sigma}_j^+e^{i\phi_j})} .
\end{equation}
In contrast to Eq.~(1) of the main text, in the form above we have retained a site-dependent phase $\phi_j$ of the drive laser at the location of the $j$th ion. However, this phase can be absorbed by a redefinition of rotated spin raising and lowering operators, $\hat{\sigma}^{\pm}_j \to e^{\pm i \phi_j}\hat{\sigma}^{\pm}_j$. Assuming phase coherence of the laser driving sideband or carrier transitions throughout the NR or SA protocols, this redefinition of the spin operators has no physical consequence for any measured observables and we may use Eq.~(1) of the main text without loss of generality. Note also, our statement in the main text that the spin-boson coupling $g_{jm}$ depends only on the participation of the $j$th ion in the $m$th normal mode implies the assumption that the intensity of the driving field is uniform across the chain.}


\section{Large N analytic model \label{sec:largeN}}
Insight into the NR and SA protocols is provided by an analytic solution of the spin-phonon dynamics that is valid in the limit of a large number of ions. In the following, we focus our discussion on the case of uniform coupling of each qubit to the CM phonon mode. 

Our analytic model is based on a Holstein-Primakoff transformation of collective spin operators. Specifically, the collective spin raising and lowering operators are replaced by boson creation and annihilation operators, $\sum_{j=1}^N \sigma^+_j \to \sqrt{N}\hat{b}^{\dagger}$ and $\sum_{j=1}^N \sigma^-_j \to \sqrt{N}\hat{b}$ \cite{HolsteinPrimakoff}. This mapping is valid for large $N$ such that $\langle \hat{b}^{\dagger}\hat{b} \rangle \ll N$ or equivalently the collective spin is strongly polarized along -$\hat{z}$. We use the Holstein-Primakoff transformation to rewrite the TC Hamiltonian [see Eq.~(1) of the main text] for the CM mode in the form of a beam-splitter Hamiltonian from quantum optics,
\begin{equation}\label{eqn:hbs}
    \hat{H}_{\mathrm{BS}} = g_0(\ha\hat{b}^{\dagger} + \ha\+\hat{b}). 
\end{equation}
We have dropped the subscripts on the phonon operators $\ha_{\mathrm{COM}} \to \hat{a}$ and $\ha\+_{\mathrm{COM}} \to \hat{a}^{\dagger}$ for brevity. Solution of the dynamics generated by the TC interaction in the Heisenberg picture yields, 
\begin{equation}
    \begin{pmatrix}
        \hat{b}(t) \\
        \hat{a}(t)
    \end{pmatrix} = \begin{pmatrix}
        \cos{(g_0 t)} && -i\sin{(g_0 t)} \\
        -i\sin{(g_0 t)}  && \cos{(g_0 t)}
    \end{pmatrix}
    \begin{pmatrix}
        \hat{b}(0) \\
        \hat{a}(0)
    \end{pmatrix}.
\end{equation}
Thus, applying the TC interaction for a duration $t = \pi/2g_0$ perfectly swaps the spin and boson lowering operators, $\hat{b}(\pi/2g_0) = -i\hat{a}(0)$ and $\hat{a}(\pi/2g_0) = -i\hat{b}(0)$. Equivalently, the spin projections proportionately map into the boson quadratures $\hat{X} = \ha\+ + \ha$ and $\hat{Y} = i(\ha\+ - \ha)$ as,
\begin{eqnarray}\label{eqn:bs-quadratures-sx}
    \hat{S}_x(\pi/2g_0) &=& \sqrt{N}\hat{Y}(0)/2, \\
    \label{eqn:bs-quadratures-sy}
    \hat{S}_y(\pi/2g_0) &=& \sqrt{N}\hat{X}(0)/2.
\end{eqnarray}

\subsection{Noise reduction protocol}
Equations (\ref{eqn:bs-quadratures-sx}) and(\ref{eqn:bs-quadratures-sy}) elucidate the underlying principle of the NR protocol: After $t = \pi/2g_0$, the fluctuations of the phonon quadratures are interchanged onto the spin observables (and vice-versa) such that $N(\Delta \hat{X})^2 /4 \leftrightarrow (\Delta\hat{S}_y)^2$ and $N(\Delta \hat{Y})^2 / 4 \leftrightarrow (\Delta\hat{S}_x)^2$. Thus, any initial squeezing of the phonon mode will be be inherited by the spin observables. Finite size effects beyond the large $N$ description become important when $\langle \hat{b}^{\dagger}\hat{b} \rangle \ll N$ is no longer satisfied throughout the TC interaction, such as when a sufficiently large amount of boson squeezing is attempted to be transferred to the spins. 



\subsection{Signal amplification protocol}
In the large $N$ limit, the imprinting of $\theta$ by a rotation of the spins in the SA protocol can be mapped to an effective enhanced displacement of the phonons due to the initial and final squeezing operations. This is seen by writing out the full expression for the spin-phonon state after the final squeezing pulse (see Fig.~2 of the main text): 
\begin{equation} \label{eqn:TimeReversalEvo}
\vert \psi_f \rangle = \underbrace{\hat{S}^{\dagger}(\zeta)\hat{U}^{\dagger}_{\mathrm{TC}}}_{\mathrm{Disentangle}} \underbrace{\hat{R}_{\varphi}^{-\frac{\pi}{2}} \hat{R}_z^{\theta} \hat{R}_{\varphi}^{\frac{\pi}{2}}}_{\mathrm{Ramsey}} \underbrace{\hat{U}_{\mathrm{TC}} \hat{S}(\zeta)}_{\mathrm{Entangle}} \vert \psi_0 \rangle,
\end{equation}
where,
\begin{equation}
\hat{R}_{\varphi}^{\frac{\pi}{2}} = \exp{\left[-i\frac{\pi}{2}(\hat{S}_x\cos{\varphi} + \hat{S}_y \sin{\varphi})\right]},
\end{equation}
and $\varphi = \phi + \frac{\pi}{2}$. The angle $\varphi$, set by $\phi$, is defined to generate an optimal rotation aligning the squeezed spin quadrature with the azimuth of the Bloch sphere during the Ramsey sequence. For simplicity, we assume $\phi = 0$ henceforth.

Making the same replacement as was done in Eq.~(\ref{eqn:hbs}) and using Eqs.~(\ref{eqn:bs-quadratures-sx}) and~(\ref{eqn:bs-quadratures-sy}), we can rewrite Eq.~(\ref{eqn:TimeReversalEvo}) as, 
\begin{eqnarray}\label{eqn:TimeReversalEff}
    \vert \psi_f \rangle & = & \underbrace{\hat{S}^{\dagger}(\zeta)}_{\mathrm{Unsqueeze}} \underbrace{\hat{U}^{\dagger}_{\mathrm{TC}} \hat{R}_{\varphi}^{-\frac{\pi}{2}} \hat{R}_z^{\theta} \hat{R}_{\varphi}^{\frac{\pi}{2}} \hat{U}_{\mathrm{TC}}}_{\mathrm{ Imprinting}} \underbrace{\hat{S}(\zeta)}_{\mathrm{Squeeze}} \vert \psi_0 \rangle \\ \notag
    & = & \hat{S}^{\dagger}(\zeta) \hat{D}(\sqrt{N}\theta/2) \hat{S}(\zeta) \vert \psi_0 \rangle .
\end{eqnarray}
where $\hat{D}(\beta) = e^{\beta\hat{a}^{\dagger}-\beta^*\hat{a}}$ is the displacement operator for the phonon mode. The rotation of the spin ensemble is thus transformed by the TC interaction into a displacement of the phonon mode. This displacement is enhanced by the squeezing according to the identity $\hat{S}^{\dagger}(\zeta) \hat{D}(\sqrt{N}\theta/2) \hat{S}(\zeta) = \hat{D}(\sqrt{N}\theta e^r/2)$ and thus the final state can be further simplified to $\vert \psi_f \rangle = \ket{\downarrow}^{\otimes N} \otimes \vert \alpha(r,\theta) \rangle_{\mathrm{ph}}$ where the phonons are characterized by a coherent state with amplitude $\alpha(r,\theta) = \sqrt{N}\theta e^r/2$. 

The large $N$ result given by Eq.~(\ref{eqn:TimeReversalEff}) provides the insight that at the end of the time-reversal sequence information about the spin rotation is encoded in the phonon mode. An optimal signal to infer $\theta$ would be the phonon quadrature $\hat{X}$ and a straightforward computation yields,
\begin{equation}
    (\Delta \theta)^2 = \frac{\langle (\Delta \hat{X} )^2 \rangle}{\vert \partial_{\theta} \langle \hat{X} \rangle \vert^2} = \frac{e^{-2r}}{N} .
\end{equation}
We identically obtain this sensitivity from a measurement of $\hat{S}_z$ by supplementing the time-reversal sequence with a final TC interaction to interchange the spin and phonon states, followed by a global $\pi/2$ spin rotation about $\hat{x}$  (see Fig.~2 of the main text). 

\subsection{Coupling to the B mode}
The results of the large $N$ model are not directly applicable to the B mode as the inhomogeneous coupling to the spin ensemble precludes the use of a similar Holstein-Primakoff transformation. However, the analytic expressions can still provide qualitative insight by first noting that a toy model of the spin-phonon coupling is to take $g_{j,\mathrm{B}} = \frac{g_0}{\sqrt{N}}\mathrm{sgn}(j - N/2)$ where $\mathrm{sgn}$ is the signum function. While this ignores the spatial variation of the magnitude of the real $g_{j,\mathrm{B}}$, it captures the $\pm$ structure. When this toy coupling is substituted into the TC Hamiltonian [Eq.~(1) of the main text] one finds all relevant physics may be described in terms of the differential operators $\hat{S}_{\alpha,-} = \frac{1}{2}\sum_{j=N/2+1}^N \hat{\sigma}^{\alpha}_j - \frac{1}{2}\sum_{j=1}^{N/2} \hat{\sigma}^{\alpha}_j$ where $\alpha = x, y$ and the collective magnetization $\hat{S}_z$, which together satisfy the usual SU(2) commutation relations. Replacing $\hat{S}_x \to \hat{S}^x_-$ and $\hat{S}_y \to \hat{S}^y_-$ throughout the prior discussion of the NR and SA protocols, all relevant results hold identically for the toy model of the B mode coupling under the assumption that $\theta$ is imprinted via a differential rotation.  

\Rev{In Figs.~1(c) and~2(b) of the main text there exists an offset between the achievable metrological gain obtained with $\hat{\mathcal{S}}_z$ and the QCRB even at moderate bosonic squeezing. To elucidate the origin of this offset, it is useful to calculate the sensitivity for the NR protocol in the extreme case with no bosonic squeezing, $\xi^2_b = 1$. For this case, the TC interaction has no effect on the spins and bosons (the phonon vacuum is dark to the sideband transition) and so the spin-boson state can be described by $\ket{\psi} = \ket{\downarrow}^{\otimes N} \otimes \ket{0}$. For the remainder of our calculation, we can thus trace away the phonons and consider only the spin degree of freedom. 
The relevant quantities to evaluate are the projection noise $\langle(\Delta\hat{\mathscr{S}}_{z})^2\rangle\approx N/4$ and the derivative of the signal $\vert \partial_{\theta} \langle \hat{\mathscr{S}}_{z}\rangle\vert^2 = N\sum_j{\vert g_{j,\mathrm{B}}\vert^2}/4g^2_0$ at the end of the Ramsey sequence assuming $\theta \to 0$.
The inhomogeneous couplings that contribute to the latter quantity must in general be numerically evaluated. However, we empirically find for large $N$ that $\vert \partial_{\theta} \langle \hat{\mathscr{S}}_{z}\rangle\vert^2 \approx N^2/(4\sqrt{2})$. The metrological gain for large $N$ is thus predicted to be, 
\begin{equation}
    N(\Delta\theta)^2 = N\frac{\langle(\Delta\hat{\mathscr{S}}_{z})^2\rangle}{\vert \partial_{\theta}\langle\hat{\mathscr{S}}_{z}\rangle\vert^2} \approx \sqrt{2} ,
\end{equation} 
which is offset above the SQL by approximately $1.5$~dB. This offset is a consequence of the fact that $\hat{\mathcal{S}}_z$ is not an optimal observable. (For the special case of $\xi^2_b = 1$ the weighted observable $\hat{S}_{z,-}$ is in fact the optimal choice to saturate the SQL). We note that an equivalent perspective is that the weighted spin operators do not obey typical SU(2) commutation relations (nor do other combinations of weighted and collective spin operators) and so, e.g., for a coherent spin state one can show (see below) $\langle (\Delta \hat{\mathscr{S}}_x)^2 \rangle \langle (\Delta \hat{\mathscr{S}}_y)^2 \rangle > \langle \hat{\mathscr{S}}_z \rangle^2$ for the weighted operators $\hat{\mathscr{S}}_{\alpha} = \frac{\sqrt{N}}{2g_0}\sum_{j=1}^N g_{j,\mathrm{B}} \hat{\sigma}^{\alpha}_j$, in contrast to the typical equality for collective spin observables.

}

\Rev{
\section{Finite size effects}

\begin{figure}
    \centering
    \includegraphics[width=8cm]{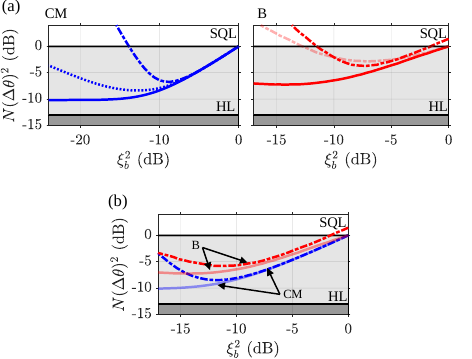}
    \caption{\Rev{Metrological gain $N(\Delta\theta)^2$ as a function of boson squeezing $\xi^2_b$ for (a) NR protocol and (b) SA protocol for a chain of $N=20$ ions. In all panels, we plot the performance achieved with measurements of $\hat{M} = \hat{S}_{z,+}$ (COM, dot-dashed line), $\hat{\mathscr{S}}_z$ (B, dot-dashed line) and $\hat{S}_{z,-}$ (B for the NR protocol, faded dot-dashed line) and compare to the QCRB given by the QFI of the full system, $N(\Delta\theta)^2 = N/F_Q$ (solid lines). In panel (a), we also compare against the spin-only QFI $F_{Q,s}$ for the NR protocol with CM mode (dotted line).}
    }
    \label{fig:SmallSystems}
\end{figure}

\subsection{Example results for larger chains}
In Fig.~\ref{fig:SmallSystems} we plot the metrological gain for the NR and SA protocols using a $1$D chain with $N=20$ ions, which is a system size that can be realized in state-of-the-art trapped ion experiments. Due to the increase in system size, we obtain the results from semiclassical numerical calculations that use the truncated Wigner approximation \cite{asier2017dtwa} (see later section). Panel (a) shows that the NR protocol can achieve an optimal gain of about $7$~dB and $4$~dB for the CM and B modes, respectively. Similarly, the SA protocol provides enhanced performance in the oversqueezed regime with an optimal gain of about $9$~dB and $6$~dB for each mode. The qualitative features of the results remain predominantly unchanged compared to the small system with $N=6$ shown in Figs.~1(c) and 2(b) of the main text. However, the increase in system size allows us to demonstrate how quickly the optimal observable for the B mode becomes $\hat{\mathscr{S}}_z$.

\subsection{Scaling of metrological gain with system size}

The behaviour of the optimal metrological gain with the number of ions is an important figure of merit that captures the scalability of our proposed protocols. In Fig.~\ref{fig:Nscaling} we show the scaling of the gain for the CM [panel (a)] and B [panel (b)] modes separately (a subset of this data for $4 \leq N \leq 50$ is shown in Fig.~2(c) of the main text). Here, we also show fits (lines) to the simulation data (markers) using the function $N(\Delta\theta)^2 = a N^{-b}$. Results for the fitting parameters are shown in Table~\ref{table:fitparam}.
}

\begin{figure}[tbh!]
    \centering
    \includegraphics[width=8cm]{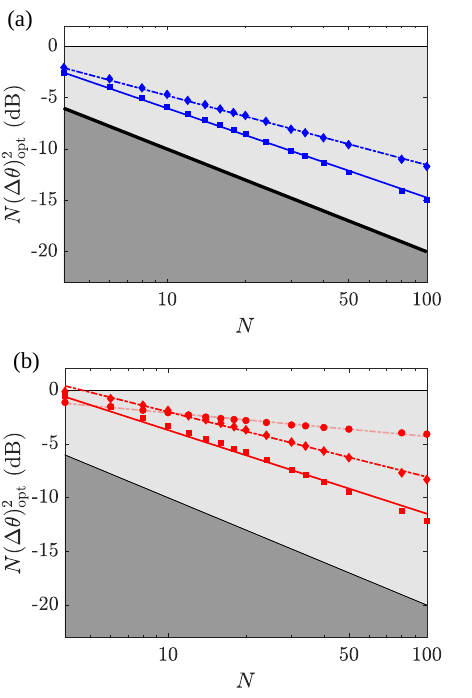}
    \caption{\Rev{Scaling of optimal metrological gain with system size for: (a) CM and (b) B mode. The results of semiclassical simulations of the NR and SA protocols are shown by markers in both panels. Diamond markers indicate the achievable gain for the NR protocol with the observable $\hat{S}_{z,+}$ and $\hat{\mathscr{S}}_z$ for the CM and B modes, respectively, while circle markers are for $\hat{S}_{z,-}$ for the B mode. Similarly, square markers indicate the achievable gain for the SA protocol with the observable $\hat{S}_{z,+}$ and $\hat{\mathscr{S}}_z$ for the CM and B modes, respectively. Fits to the simulation data (see text) are indicated by lines: dot-dashed and solid lines for the NR and SA protocols, respectively. We distinguish $\hat{S}_{z,-}$ (faded) and $\hat{\mathscr{S}}_z$ (darker) measurements for the NR protocol with the B mode.}
    }
    \label{fig:Nscaling}
\end{figure}

\begin{table}[tbh!]
\begin{ruledtabular}
\begin{tabular}{cccc}
Protocol & Mode  & $a$ & $b$ \\
\hline
NR & CM ($\hat{S}_{z,+}$) & $0.45\pm0.05$ & $0.68\pm0.02$ \\
NR & B ($\hat{\mathscr{S}}_{z}$) & $0.93\pm0.08$ & $0.61\pm0.03$ \\
NR & B ($\hat{S}_{z,-}$) & $0.02\pm0.02$ & $0.22\pm0.01$ \\
SA & CM ($\hat{S}_{z,+}$) & $0.62\pm0.07$ & $0.87\pm0.03$ \\
SA & B ($\hat{\mathscr{S}}_{z}$) & $0.9\pm0.2$ & $0.77\pm0.06$ 
\end{tabular}
\end{ruledtabular}
\caption{\label{table:fitparam} \Rev{Summary of fitted parameters for each protocol and mode shown in Fig.~\ref{fig:Nscaling}, including uncertainty due to fitting error within a $95$\% confidence interval. Measured observable used to compute the metrological gain is indicated in brackets. Note that the fit for the NR protocol using the B mode and measuring $\hat{S}_{z,-}$ is poorly conditioned.} }
\end{table}

\section{Role of spin-phonon entanglement}
The performance of the NR protocol is limited by the build-up of spin-phonon entanglement as a result of finite size effects neglected by the large $N$ model. The entanglement strongly distinguishes our scheme from the analogous oversqueezing observed in related collective spin systems that generate spin squeezing using twisting dynamics \cite{Kitagawa1993}. In the latter case, it is well established that oversqueezing, which occurs when fluctuations of the quantum state wrap around the collective Bloch sphere, leads to poor sensitivity if the mean spin projection is used as the measurement signal to infer $\theta$. However, measurements of higher-order observables \cite{gessner2022nonlinear}, distribution functions \cite{Strobel2014,Bohnet2016} or even parity \cite{bollinger1996optimal} are sufficient to saturate the QCRB and obtain performance near the Heisenberg limit.

\begin{figure}
    \centering
    \includegraphics[width=8cm]{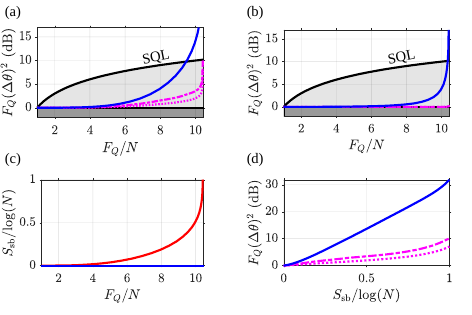}
    \caption{(a) Normalized metrological gain $F_Q(\Delta\theta)^2$ as a function of QFI per particle $F_Q/N$ for the NR protocol. The gain is computed using a measurement of $\hat{S}_{z,+}$ (blue solid line, see main text), and the QFI $(\Delta\theta)^2 = 1/F_{Q,s}$ (magenta dotted line) and CFI $(\Delta\theta)^2 = 1/F_{C,s}$ (magenta dot-dashed line) of the spin subsystem. The light shaded region indicates sub-SQL performance for reference (the forbidden region below the QCRB [$F_Q(\Delta\theta)^2 = 1$] is also indicated by dark shading). (b) Same but for the SA protocol. (c) Spin-boson entanglement entropy $S_{\mathrm{sb}}$ for the NR protocol (red solid line) and SA protocol (blue solid line). In the former case, the entanglement is computed for the input state $\hat{\rho}_{\mathrm{in}}$, while the latter is computed at the end of the SA protocol. (d) Parametric plot of the metrological gain $F_Q(\Delta\theta)^2$ versus entanglement entropy $S_{\mathrm{sb}}$ for the NR protocol. Lines follow the same convention as panel (a).   }
    \label{fig:SpinPhEntang}
\end{figure}

In contrast, Fig.~\ref{fig:SpinPhEntang}(a) demonstrates that measurements of the spin degree of freedom are insufficient to saturate the QCRB with the NR protocol. We compute the metrological gain normalized by the QFI (i.e., relative to the QCRB), $F_Q(\Delta \theta)^2$, and plot it as a function of the QFI per particle, $F_Q/N = 4\langle (\Delta \hat{S}_{z,+})^2 \rangle/N$ for an example system of $N=20$ ions that are coupled to the CM mode. The blue solid line indicates the gain computed with a measurement of $\hat{S}_{z,+}$ for the NR protocol. When the QFI per particle is relatively small ($F_Q/N \lesssim 4$), we find $F_Q(\Delta \theta)^2 \simeq 1$ and thus the measurement is sufficient to nearly saturate the QCRB (defined by $F_Q(\Delta \theta)^2 = 1$). On the other hand, when the QFI per particle grows larger (at most, $F_Q/N \simeq N^2/2 = 10$ in this instance), $F_Q(\Delta \theta)^2$ rapidly grows indicating an achievable sensitivity far worse than the QCRB. 

Similar behaviour is found when the gain is computed from either the quantum or classical Fisher information (CFI) of the spin subsystem, equivalent to $F_Q(\Delta \theta)^2 = F_Q/F_{Q,s}$ and $F_Q/F_{C,s}$, respectively. The QFI is obtained using,
\begin{equation}
    F_{Q,s} = 4 \lim_{\theta \to 0} \frac{1 - \mathcal{F}(\rho_{\mathrm{in},s},\rho_{\theta,s})}{\theta^2} ,
\end{equation}
where 
\begin{equation}
    \mathcal{F}(\hat{\rho}_{\mathrm{in},s},\hat{\rho}_{\theta,s}) = \left( \mathrm{Tr}_s\left[ \sqrt{ \sqrt{\hat{\rho}_{\mathrm{in},s}} \hat{\rho}_{\theta,s} \sqrt{\hat{\rho}_{\mathrm{in},s}} } \right] \right)^2 ,
\end{equation}
is the Uhlmann fidelity between the reduced density matrices of the spin degree of freedom of the input (probe) state 
$\hat{\rho}_{\mathrm{in},s} = \mathrm{Tr}_{\mathrm{ph}}\left[ \hat{\rho}_{\mathrm{in}} \right]$ and the output state $\hat{\rho}_{\theta,s} = \mathrm{Tr}_{\mathrm{ph}}\left[ \hat{R}^{\theta}_{z,+} \hat{\rho}_{\mathrm{in}} \hat{R}^{-\theta}_{z,+} \right]$. Here, the input state is taken to be that right before $\theta$ is imprinted by the spin rotation, 
\begin{equation}
    \hat{\rho}_{\mathrm{in}} = \hat{R}^{\frac{\pi}{2}}_{y}\hat{U}_{\mathrm{TC}} \hat{S}(\zeta) \vert \psi_{0} \rangle \langle \psi_{0} \vert \hat{S}^{\dagger}(\zeta) \hat{U}^{\dagger}_{\mathrm{TC}} \hat{R}^{-\frac{\pi}{2}}_{y} .
\end{equation}
The CFI is obtained from, 
\begin{equation}
    F_{C,s}(\theta) = \sum_{m_{\mathbf{n}}} \frac{1}{P(m_{\mathbf{n}}\vert \theta )} \left( \frac{d P(m_{\mathbf{n}}\vert \theta )}{d\theta} \right)^2 ,
\end{equation}
where $P(m_{\mathbf{n}}\vert \theta ) = \mathrm{Tr}[\vert m_{\mathbf{n}} \rangle \langle m_{\mathbf{n}} \vert \hat{\rho}_{\theta,s}]$ and $\vert m_{\mathbf{n}} \rangle$ is an eigenstate of the collective spin projection $\hat{S}_{\mathbf{n}}$ along some unit vector $\mathbf{n}$. Without loss of generality, we work around $\theta = 0$ and optimize over $\mathbf{n}$ in all figures. The gains computed from the CFI and QFI of the spin subsystem, shown in Fig.~\ref{fig:SpinPhEntang}(a), follow a similar trend to that obtained from $\hat{S}_{z,+}$. However, they additionally emphasize that it is not possible to saturate the QCRB with spin measurements when the QFI of the full spin-boson system approaches relatively large values (i.e., on the order of $N^2$). In fact, for $F_Q/N \approx N^2/2$ the CFI indicates that a measurement of a spin projection yields performance worse than the SQL. 

The loss of metrological gain in the NR protocol is a consequence of the build-up of appreciable spin-boson entanglement. This is demonstrated by plotting the Renyi entanglement entropy $S_{\mathrm{sb}} = -\mathrm{log}(\mathrm{Tr}[\hat{\rho}_{\mathrm{in},s}^2])$ as a function of the QFI per particle in Fig.~\ref{fig:SpinPhEntang}(c). The entanglement increases nonlinearly with the QFI, i.e., the potential metrological utility of the underlying quantum state. Noticeably, the entanglement first becomes appreciably non-zero for $F_Q/N \gtrsim 4$, which correlates with the oversqueezed regime discussed above, and appears to qualitatively mirror the behaviour of $F_Q(\Delta \theta)^2$ shown in panel (a) (note the logarithmic scale of the vertical axis in the latter). This observation can be elegantly explained by the fact that the Renyi entropy quantifies the mixedness of the reduced subsystem, $\mathrm{Tr}[\hat{\rho}_{\mathrm{in,s}}^2]$. Entanglement between the spin and phonon degrees of freedom leads to mixedness and thus excess quantum noise in the spin subsystem after the phonons are traced away. In turn, this excess noise constrains how precisely $\theta$ can be inferred from spin measurements. Strikingly, a parametric plot of the normalized metrological gain $F_Q(\Delta \theta)^2$ and the entanglement entropy in Fig.~\ref{fig:SpinPhEntang}(d) shows that the former is approximately exponentially dependent on the latter. Thus, one can actually only saturate the QCRB [$F_Q(\Delta\theta)^2 = 1$] with access to the complete spin-phonon state or when the two degrees of freedom are disentangled.

To complete our discussion, we also plot the normalized metrological gain as a function of the QFI per particle for the SA protocol in Fig.~\ref{fig:SpinPhEntang}(b). The blue solid line indicates the gain computed with a measurement of $\hat{S}_{z,+}$, which for the SA protocol stays closer to the QCRB for a larger range of QFI per particle but still diverges for $F_Q/N \sim N^2/2$. On the other hand, the fact that the final state at the end of the SA protocol is a disentangled product of spins and phonons [see the vanishing Renyi entropy in Fig.~\ref{fig:SpinPhEntang}(c)] leads to the gain computed from the QFI and CFI of the spin subsystem saturating the QCRB across the whole range of $F_Q/N$.

\section{Technical noise and finite temperature effects}
In the main text we comment on the impact of a number of sources of technical noise and decoherence. Here, we provide further detail and supporting calculations. Our results are tailored to the CM mode, but are found to also qualitatively explain numerical results for the B mode (see Fig.~3 of the main text). 

\subsection{Thermal noise}
An initial thermal occupation $\bar{n}$ of the phonon mode leads to excess projection noise that degrades the performance of both NR and SA protocols in a similar fashion. However, we treat each protocol separately. 

We first consider the NR protocol. In the limit of infinite $N$, we simply have that the spin squeezing will scale as $\xi^2_s = (2\bar{n}+1)e^{-2r}$ as a result of the fact that the phonons are initially described by a squeezed thermal state with excess (i.e., larger than vacuum) quadrature fluctuations $\langle (\Delta\hat{Y})^2 \rangle = (2\bar{n}+1) e^{-2r}$ and $\langle (\Delta\hat{X})^2 \rangle = (2\bar{n}+1)^2 e^{2r}$. However, this excess projection noise of the initial boson state has additional effects when finite $N$ is considered. Specifically, the limit on the squeezing that can be mapped to the spin degree of freedom is intimately related to the scale of the anti-squeezed spin quadrature relative to the scale of the curvature of the Bloch sphere. From the ideal scaling of spin-squeezing with $N$ that we present in Fig.~2 of the main text, we can assume that the degree of anti-squeezing that can be supported before curvature effects become important is $\langle (\Delta \hat{S}_{x,+})^2\rangle/N \sim N^{2/3}$ for large $N$. Then, we equate this to the fluctuations of the bosonic squeezed state that we couple the spins to, $N^{2/3} \sim e^{2r}(2\bar{n}+1)$. Rearranging and solving for the squeezing strength gives an optimal $r_{\mathrm{opt}} = \frac{1}{2}\log[N^{2/3}/(2\bar{n}+1)]$ and thus the best possible spin squeezing after transfer is $\xi^2_{s,\mathrm{opt}} \simeq e^{-2r_{\mathrm{opt}}}(2\bar{n}+1) = (2\bar{n}+1)^2/N^{2/3}$. Thus, thermal noise reduces the best achievable spin squeezing by a factor of $(2\bar{n}+1)^2$. 

\begin{figure}
    \centering
    \includegraphics[width=8cm]{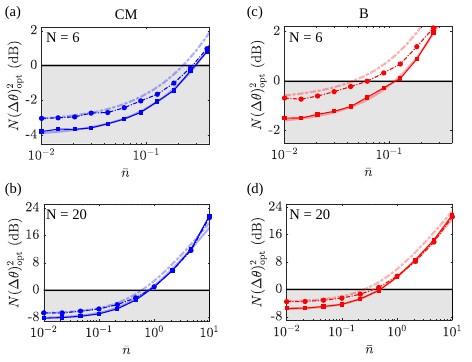}
    \caption{\Rev{Optimal metrological gain as a function of initial thermal occupation $\bar{n}$ of the: (a)-(b) CM and (c)-(d) B mode. In all panels we plot numerical results from a semiclassical calculation for the NR (connected circles) and SA (connected squares) protocols. Motivated by the large $N$ theory, we compare the semiclassical results to semi-analytic expressions (faded dot-dashed and solid lines for NR and SA protocols, respectively) obtained by scaling the $\bar{n}=0$ results with a correcting factor $(2\bar{n}+1)^2$. For the B mode, we distinguish results for the NR protocol using final measurements of $\hat{S}_{z,-}$ (faded circles) and $\hat{\mathscr{S}}_z$ (dark circles). In both panels the shaded grey region indicates quantum-enhancement. The system size is indicated in each panel.}}
    \label{fig:ThermalNoise}
\end{figure}

\Rev{
We compare our analytic prediction for the NR protocol to numerical results obtained with finite $N$ in Figs.~\ref{fig:ThermalNoise}(a) and (b). The $(2\bar{n}+1)^2$ scaling of the optimal gain is reasonably replicated for $\bar{n} \ll 1$, even in the limit of a small system with $N=6$. Quantitative discrepancies are attributed to the fact that our large $N$ theory is based on a heuristic treatment of the leading order finite size effects. Concretely, our numerical results indicate that the NR protocol is largely unaffected by thermal effects for $\bar{n} \lesssim 0.1$ (and this robustness improves with system size), which is achievable in state-of-the-art trapped-ion experiments. 
}

We next consider the SA protocol and start by taking the large $N$ limit to gain intuition. In this case, the performance of the Tavis-Cummings interaction is not affected by the thermal noise and from Eq.~(\ref{eqn:TimeReversalEff}) we deduce that the sensitivity is degraded by a factor $(2\bar{n}+1)$ due to the initial excess noise of the bosons. However, numerical results with finite $N$, shown in  \Rev{Figs.~\ref{fig:ThermalNoise}(a) and (b)}, indicate that thermal fluctuations rescale the optimal sensitivity by a factor $\sim (2\bar{n}+1)^2$. Collectively, we understand this result by comparing to the NR protocol: One factor of $(2\bar{n}+1)$ comes directly from the excess noise of the initial boson state, while a second factor of $(2\bar{n}+1)$ arises due to finite size corrections. 

\Rev{
We also plot numerical results for the NR and SA protocols using the B mode in Figs.~\ref{fig:ThermalNoise}(c) and (d). We observe a similar reduction in performance with added thermal noise, with the achievable gain approximately rescaled by a factor $(2\bar{n}+1)^2$. 
}
This demonstrates that the approximate analytic treatment developed above can also provide intuition for the effects of thermal noise on protocols involving the B mode. For clarity, we also point out that thermal noise will identically affect the SA protocol regardless of whether $\hat{\mathscr{S}}_z$ or $\hat{S}_{z,+}$ is measured. Note that in the latter case, the CM mode, which is coupled to for readout, may also feature a thermal occupation. However, the efficacy of the \emph{bosonic} beamsplitter implemented between the B and CM modes is state-independent and thus does not limit the sensitivity.

\subsection{Phase fluctuations}
A lack of phase stability between the Tavis-Cummings and boson squeezing operations can be incorporated into our analysis by allowing the squeezing phase to fluctuate \cite{FossFeig2019amplification,Burd2023squeezing}. To be concrete, we adopt a model wherein the squeezing phase $\phi$ is treated as a normal random variable with mean $\overline{\phi} = 0$ and variance $\overline{\phi^2} = \sigma^2$. Here, the overline indicates a classical average over the phase noise. Our model assumes that the squeezing phase is fixed across a single experimental trial, which is consistent with the fact that the phase of each operation, e.g., the TC interaction or boson squeezing/beam-splitter operations, would ideally be individually stable across a single trial. In the following, we present separate treatments of the NR and SA protocols with phase fluctuations in the large $N$ limit. 

For the NR protocol, fluctuations of the phase $\phi$ wash out the squeezing of the initial bosonic state~\footnote{Note that, alternatively, one could understand the fluctuations as leading to a random orientation of the spin squeezed state if the relative phase of the sideband is instead varied.}. As the the fluctuations of the phonon mode and spin ensemble are interchanged perfectly by the TC interaction in the large $N$ limit, the reduction of boson squeezing directly leads to a reduction in the level of spin squeezing that can be achieved. With this in mind, the metrological gain of the NR protocol (equivalent to spin squeezing) is given by, 
\begin{multline}
    N(\Delta\theta)^2 \equiv \overline{\langle (\Delta \hat{X})^2 \rangle} = \frac{e^{-2r}}{2}\left( 1 + e^{-2\sigma^2}\right) \\+ \frac{e^{2r}}{2}\left( 1 - e^{-2\sigma^2}\right) .
\end{multline}
The last term of this expression elucidates that fluctuations of the squeezing phase $\phi$ about the optimal value of zero introduces a deleterious contribution from the anti-squeezed quadrature. A key result is that we find there exists an optimal boson squeezing,
\begin{equation}
    r_{\mathrm{opt}} = \frac{1}{2}\log\left( \frac{\sqrt{1-\sigma^2}}{\sigma} \right)
\end{equation}
that is set by the strength of the phase fluctuations and leads to the optimal metrological gain, 
\begin{equation}\label{eqn:PhaseNoise_NR}
   N(\Delta\theta)^2_{\mathrm{opt}} = \sqrt{(1 -e^{-2\sigma^2})(1+e^{-2\sigma^2})} .
\end{equation}
For $\sigma \ll 1$ this simplifies to $N(\Delta\theta)^2_{\mathrm{opt}} \approx 2\sigma$. It is useful to identify that for $\sigma \gtrsim N^{-2/3}$, phase fluctuations become the limiting factor for the performance of the NR protocol, rather than finite size effects.  

The effect of phase fluctuations on the SA protocol is best understood in light of the previous analysis presented in the large $N$ analytic model. However, a subtle difference is that the time-reversal sequence simplifies to,
\begin{multline}
    \hat{S}(-re^{2i\phi}) \hat{U}^{\dagger}_{\mathrm{TC}} \hat{R}_{\varphi}^{-\frac{\pi}{2}} \hat{R}_z^{\theta} \hat{R}_{\varphi}^{\frac{\pi}{2}} \hat{U}_{\mathrm{TC}} \hat{S}(re^{2i\phi}) \\ = \hat{D}\left(\frac{\sqrt{N}\theta\cosh(r) + e^{2i\phi}\sqrt{N}\theta\sinh(r)}{2}\right) .
\end{multline}
Note that time-reversal in this sequence is realized by jumping the squeezing phase $\phi \to \phi + \pi/2$. Following the same analysis as previous and averaging over the squeezing phase we obtain the metrological gain (see main text for relevant definitions for the SA protocol),
\begin{equation}\label{eqn:PhaseNoise_SA}
    N(\Delta\theta)^2 = \left[ \cosh(r) + e^{-2\sigma^2}\sinh(r) \right]^{-2} .
\end{equation}
In the limit of strong squeezing but weak phase noise, $e^{-r} \ll 1$ and $\sigma\ll 1$, this result yields the perturbative form $N(\Delta\theta)^2 \approx e^{-2r}(1 + 2\sigma^2)$. On the other hand for strong phase noise $\sigma \gg 1$ we observe that Eq.~(\ref{eqn:PhaseNoise_SA}) collapses to $N(\Delta\theta)^2 \approx [\cosh^{-2}{r}]$, which at worst predicts a four-fold increase (i.e., loss of $6$~dB of metrological gain) over the ideal ($\sigma = 0$) result.

\begin{figure}
    \centering
    \includegraphics[width=8cm]{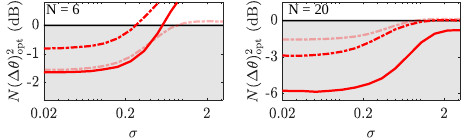}
    \caption{Optimal metrological gain as a function of phase noise magnitude $\sigma$ for NR (dot-dashed lines) and SA (solid lines) protocols \Rev{with $N=6$ and $N=20$} for B mode. Faded and solid dot-dashed lines distinguish $\hat{M} = \hat{S}_{z,-}$ and $\hat{\mathscr{S}}_z$, respectively.  
    }
    \label{fig:Noise_SM}
\end{figure}

Figure 3 of the main text shows numerical results for the NR and SA protocols using the CM mode including the effects of phase fluctuations for a pair of systems with $N = 6$ and $N=20$. The results in that instance demonstrate that the large $N$ analytic predictions provide useful insight even in the limit of small system size. Similarly, Fig.~\ref{fig:Noise_SM} shows numerical results for protocols using the B mode for $N=20$ and $N=6$ (both are obtained from semiclassical calculations using the truncated Wigner approximation). While the analytic predictions are no longer directly applicable, we find that for sufficiently large systems the qualitative predictions are still relevant. In particular, both protocols are robust to perturbative phase noise $\sigma \ll 1$ and in the limit of large phase noise $\sigma \gg 1$ the SA protocol retains some degree of quantum enhancement.

\subsection{Spin decoherence}
As discussed in the main text, single-particle decoherence of the qubits at rate $\Gamma$ will generically degrade any spin squeezing or entanglement generated during the NR or SA protocols and thus limit their metrological performance. Here, we provide further elaboration on two relevant considerations. 

First, it is useful to make a comparison of the typical timescales of the NR and SA protocols, which leverage the spin-boson TC interaction, with previously investigated approaches to generate squeezing via one-axis twisting (OAT) realized by boson-mediated collective spin-spin interactions. Our protocols are dominated by the timescale over which it takes to exchange the spin and boson states, i.e., $t_{\pi} = \pi/2g_0$~\footnote{We assume that the time required to appreciably squeeze or unsqueeze the bosons is negligible.}. Notice that this timescale is independent of the number of qubits $N$ [see Eq.~(\ref{eqn:hbs})]. On the other hand, the characteristic strength of boson-mediated spin-spin interactions is given by, e.g., $\chi \sim g_0^2/(\delta N)$ where $\delta$ is the detuning from the CM mode of a pair of applied optical dipole force beams that couple the spin and motion \cite{Britton2012,Bohnet2016}. Optimal spin squeezing (i.e., $\xi^2_s \sim N^{-2/3}$) occurs at $\chi t_{\mathrm{OAT}} \sim N^{-2/3}$ (see SM of Ref.~\cite{RLS_TSS_2018}) or equivalently $t_{\mathrm{OAT}} \sim N^{1/3}$, illustrating that squeezing via OAT becomes slower with increasing system size. As $\Gamma$ is independent of $N$, this implies that our NR and SA protocols built on the resonant TC interaction may scale favorably relative to decoherence, which can be particularly relevant for $2$D and $3$D arrays \cite{Bohnet2016,Roos_2Dcrystal_2023}.  

A second aspect is that $\Gamma$ is typically dominated by off-resonant Raman scattering in state-of-the-art trapped ion experiments \cite{bollinger2023spontaneousemission}. This scattering can be reduced by increasing the detuning $\Delta$ of the Raman beams from excited states as $\Gamma \propto 1/\Delta^2$. Assuming a commensurate increase in laser power, the spin-boson interaction $g_0\propto 1/\Delta$ can be maintained at a constant level (i.e., the coherent timescales are unaffected) while $\Gamma/g_0 \propto 1/\Delta$, i.e., the ratio of coherent and incoherent timescales, is suppressed. We note that this approach does not work for OAT protocols as the coherent dynamics is instead driven by $\chi \propto g_0^2$ and thus $\Gamma/\chi$ is independent of the detuning $\Delta$.

\section{Semiclassical calculations}
\begin{figure*}[ht]
    \centering
    \includegraphics[width=16cm]{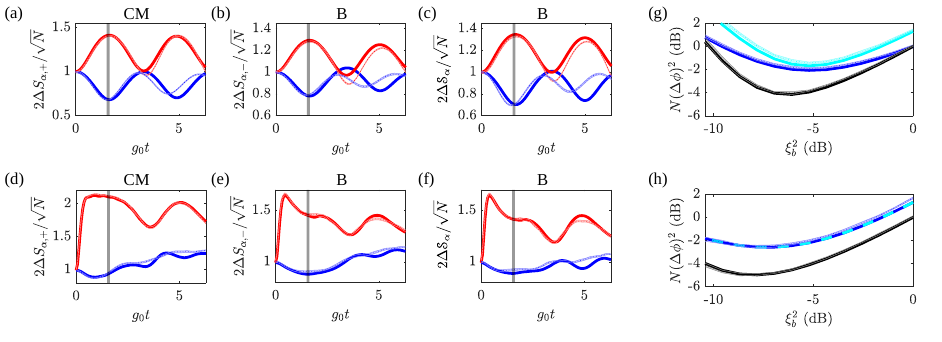}
    \caption{(a)-(f) Evolution of spin quadrature variances under a TC interaction obtained from semiclassical (faded dotted lines) and exact diagonalization (solid lines) calculations. The width of the lines correspond to the sampling error for the semiclassical calculations, which are obtained from $10^4$ trajectories. We compare spin fluctuations along $\hat{x}$ (red data) and $\hat{y}$ (blue data). Panels (a)-(c) are for squeezing strength $r = 0.4$ and panels (d)-(f) are for $r = 0.8$. Coupling to the CM or B mode is indicated. (g) Metrological gain as a function of initial boson squeezing for NR protocol obtained from semiclassical (markers) and exact diagonalization (solid lines) calculations. We compare for coupling to the CM mode (black data) and B mode (blue data: obtained with $\hat{S}_{z,-}$ and cyan data: obtained with $\hat{\mathscr{S}}_z$) separately. (h) Metrological gain as a function of initial boson squeezing for SA protocol obtained from semiclassical (faded dotted lines) and exact diagonalization (lines) calculations. The width of the lines correspond to the sampling error for the semiclassical calculations, which are obtained from $10^5$ trajectories. We compare for coupling to the CM mode (black data) and B mode (blue data: obtained with $\hat{S}_{z,+}$ and cyan dashed line: obtained with $\hat{\mathscr{S}}_z$) using exact diagonalization. All panels in this figure use $N=8$.}
    \label{fig:BenchmarkDTWA}
\end{figure*}

In various parts of the main text and this Supplemental Material, we use semiclassical calculations based on the truncated Wigner approximation (TWA). This enables us to efficiently simulate large systems and to treat the inhomogeneous coupling of the B mode to the spin ensemble. A description of the method for systems featuring both discrete (spin) and continuous (boson) degrees of freedom is outlined in Ref.~\cite{asier2017dtwa}. 

As TWA is an approximate method, we present some example benchmark results in Fig.~\ref{fig:BenchmarkDTWA} comparing to exact diagonalization for a small system of $N = 8$ spins (with a truncated Fock basis to capture the boson dynamics). Sampling error for the TWA results is indicated by the width of the plotted lines (see below), with $10^4$ trajectories used in panels (a)-(f) and $10^5$ trajectories in panels (g) and (h). The TWA data plotted in the main text (i.e., all results for coupling to the B mode, and all results in Figs.~2(c) and 3) typically use between $10^4$ and $10^5$ trajectories, depending on the calculated observable, and the sampling error (not shown in the main text) is on the order of the linewidth or marker size, which we consider numerically converged. 

Figures \ref{fig:BenchmarkDTWA}(a)-(f) plot the evolution of the spin quadratures for different amounts of initial boson squeezing and coupling to different phonon modes. We observe that the semiclassical calculation quantitatively captures the exchange of fluctuations between the spins and bosons for $t \lesssim \pi/2g_0$ for both small and large amounts of squeezing [panels (a)-(c) are for $r=0.4$ and panels (d)-(f) are for $r=\mathrm{log} N$ for reference]. At longer times, the agreement between the methods degrades, but we highlight that this corresponds to a regime that is not relevant to our NR or SA protocols. Although not shown in this figure, we also find that the agreement between the methods tends to improve as the number of spins is increased. 

Figures \ref{fig:BenchmarkDTWA}(g) and (h) additionally show predictions for the best sensitivity for the NR and SA protocols as a function of initial boson squeezing for each method. The semiclassical and exact calculations closely follow each other, with only small disagreement. In fact, it is noteworthy that the semiclassical calculation tends to fractionally \emph{underestimate} the metrological gain for the B mode.

\end{document}